\documentclass[%
 twocolumn,
 superscriptaddress,
 preprintnumbers,
 nofootinbib,
 amsmath,amssymb,
 aps,
 prl
 ]{revtex4-2}
\usepackage{slashed,bm}
\usepackage{url}
\usepackage{graphicx}
\usepackage[colorlinks = true,
            linkcolor = blue,
            urlcolor  = blue,
            citecolor = blue,
            anchorcolor = blue]{hyperref}

\allowdisplaybreaks

\begin{document}


\title{Four-loop beta function in $\mathbf{\mathcal{N}=2}$ supersymmetric Yang--Mills theory, dimensional reduction, and treatment of $\mathbf{\gamma_5}$}

\author{Maxim A.~Bezuglov}
\email{maxim.bezuglov@desy.de}
\affiliation{II. Institute for Theoretical Physics, Hamburg University,
  Luruper Chaussee 149, 22761 Hamburg, Germany}
\author{Bernd A.~Kniehl}
\email{kniehl@desy.de}
\affiliation{II. Institute for Theoretical Physics, Hamburg University,
  Luruper Chaussee 149, 22761 Hamburg, Germany}
\author{Vitaly N.~Velizhanin}
\email{vitaly.velizhanin@desy.de}
\affiliation{II. Institute for Theoretical Physics, Hamburg University,
  Luruper Chaussee 149, 22761 Hamburg, Germany}


\begin{abstract}
Adopting the dimensional-reduction scheme, we diagrammatically compute, at four loops, the beta function of $\mathcal{N}=2$ supersymmetric Yang--Mills theory, which is well known to vanish beyond one loop.
This provides a formidable testbed for both the treatment of the problematic $\gamma_5$ Dirac matrix in $D$ dimensions and the implementation of the dimensional-reduction scheme.
In the use of the reading-point method of $\gamma_5$, being preferable for convenience, specific details of implementation start to matter at four loops.
We thus rigorously validate these implementation issues, which have previously been advocated for the calculation of the gauge-coupling beta functions at four loops in the Standard Model.
\end{abstract}


\maketitle

Dimensional regularization (DREG)~\cite{Bollini:1972ui,tHooft:1972tcz} has become the standard framework for handling radiative corrections in the Standard Model (SM) and other quantum field theories of particle and mathematical physics.
One of the main advantages of DREG is that it preserves the essential symmetries of gauge theories.
In the SM, Lorentz covariance is maintained upon continuation to $D=4-2\epsilon$ dimensions, while the gauge structure of the theory is preserved at the quantum level through Becchi--Rouet--Stora--Tyutin invariance \cite{Becchi:1974xu,Becchi:1974md,Becchi:1975nq,Tyutin:1975qk} and the associated Slavnov--Taylor identities \cite{Taylor:1971ff,Slavnov:1972fg}.
As a result, the ultraviolet divergences can be renormalized within the usual gauge-invariant framework, without the need for additional finite counterterms to restore the gauge symmetry.
These properties, together with the straightforward power counting and the applicability of the method to multiloop calculations, have made DREG the standard tool for perturbative calculations in the SM.

Despite these advantages, DREG faces two well-known challenges: its compatibility with supersymmetry and the consistent treatment of $\gamma_5$ in theories with chiral fermions, including the SM. 
Supersymmetry relies on relations between bosonic and fermionic degrees of freedom that are specific to four dimensions, making a straightforward continuation to $D$ dimensions problematic. 
The treatment of $\gamma_5$ is similarly problematic, since its four-dimensional definition and anticommutation relation with other Dirac matrices cannot be consistently maintained in arbitrary $D$ dimensions.

The dimensional-reduction scheme (DRED) \cite{Siegel:1979wq} was proposed as a convenient regularization method for supersymmetric theories, as it preserves the equality between bosonic and fermionic degrees of freedom required by supersymmetry, which is otherwise violated in DREG \cite{Bollini:1972ui,tHooft:1972tcz}. 
As is well known, the original formulation \cite{Siegel:1979wq} is inconsistent beyond low orders \cite{Siegel:1980qs}, and various modifications have been proposed to solve this problem \cite{Avdeev:1981vf,Avdeev:1982xy,vanDamme:1984ig,Stockinger:2005gx}.
In practice, the original formulation of DRED may be rescued by accommodating $\epsilon$-scalars \cite{Capper:1979ns,Nicolai:1980km,Avdeev:1980bh} to overcome the split-up between four- and $D$-dimensional objects.    

As for the $\gamma_5$ problem, commonly used prescriptions include the Breitenlohner--Maison--'t~Hooft--Veltman (BMHV) scheme \cite{tHooft:1972tcz,Breitenlohner:1977hr}, Larin's prescription \cite{Larin:1993tq}, and the reading-point method \cite{Korner:1991sx}.
The BMHV scheme provides a mathematically rigorous formulation of $\gamma_5$ in DREG by treating it as a purely four-dimensional object, so that $\gamma_5$ anticommutes with the four-dimensional Dirac matrices but commutes with their $(D-4)$-dimensional components.
Larin's prescription treats $\gamma_5$ in DREG using its four-dimensional Levi-Civita-tensor representation and introduces a finite renormalization to restore its anticommutation with $D$-dimensional Dirac matrices.
The reading-point method specifies an appropriate starting point for the evaluation of fermionic traces involving $\gamma_5$, thereby fixing the ordering of the Dirac matrices when the cyclicity of the trace is not preserved in DREG.

In practice, the reading-point method is preferable for the technical convenience of its application, while the BMHV scheme quickly becomes prohibitively complicated as one proceeds to higher loop orders and Larin's prescription often only provides partial solutions. 
However, it is crucial to notice that there is a certain degree of freedom in the implementation of the reading-point method at higher loop orders, beyond the prescription originally outlined in Ref.~\cite{Korner:1991sx}.
In fact, additional rules have to be judiciously devised to guarantee the correctness of the outcome.

For certain quantities, such as gauge-coupling beta functions, the $\gamma_5$ issue can be avoided by algebraic methods based on the Weyl consistency condition (WCC) \cite{Jack:2013sha,Antipin:2013sga,Poole:2019kcm}, which allows all beta-function coefficients to be expressed in terms of a few coefficients that are not affected by $\gamma_5$ \cite{Poole:2019txl,Poole:2019kcm,Davies:2019onf,Bednyakov:2021qxa}.
But this often requires additional calculations.

In our previous work \cite{Bezuglov:2026okb}, we extended the reading-point prescription from Ref.~\cite{Bednyakov:2015ooa}, where electroweak corrections to the strong-coupling beta function in the gaugeless limit were computed, to the full SM and found agreement with the WCC approach \cite{Davies:2019onf,Bednyakov:2021qxa}.
Together with our earlier calculations in extended supersymmetric Yang--Mills (SYM) theories \cite{Velizhanin:2008rw,Velizhanin:2010vw,Kniehl:2023bbk}, this provides the basis for the present computation of four-loop beta functions in supersymmetric models.

It is worth noting that, while there is only one coupling constant in each SYM theory, its beta function can be determined from different vertices.
Supersymmetry ensures that the beta functions calculated from different vertices are the same.
Moreover, for $\mathcal{N}=2,4$ SYM theories, the beta functions are known exactly to all orders in perturbation theory \cite{Mandelstam:1982cb,Brink:1982wv,Howe:1983wj}.
This makes them an ideal testing ground for the applicability of DRED and the consistency of the $\gamma_5$ prescription in the relevant Feynman diagrams.
As suggested by our explicit four-loop analysis below, a nontrivial test of the $\gamma_5$ prescription is only feasible in $\mathcal{N}=2$ SYM theory, while the ``na\"{\i}ve'' $\gamma_5$ prescription does work for $\mathcal{N}=1,4$.

Following the successful four-loop calculations in the SM \cite{Bednyakov:2015ooa,Bezuglov:2026okb}, we now turn our attention to the analogous calculation in the $\mathcal{N}=1,2,4$ SYM theories.
As in Refs.~\cite{Bednyakov:2015ooa,Bezuglov:2026okb}, we take advantage of the
background-field method~\cite{Abbott:1980hw,Abbott:1981ke}.
Since the latter introduces additional vertices only in the pure gauge sector, the Feynman rules for the respective QCD vertices readily carry over to the SYM theories.
The analysis in $\mathcal{N}=1,2,4$ SYM is similar to the SM case, except that it proceeds in DRED, rather than DREG.

Considering a generic one-particle-irreducible Green's function $\Gamma$ at some kinematic invariant $Q^2$, we may calculate its renormalization constant $Z_\Gamma$ by exploiting its multiplicative renormalizability following Ref.~\cite{Larin:1993tp} (see also Refs.~\cite{Tarasov:1976ef,Vladimirov:1979zm,Tarasov:1980nsq}).
Specifically, $Z_\Gamma$ relates the bare (B), dimensionally regularized version of $\Gamma$ with its renormalized counterpart as
\begin{eqnarray}
\label{multren}
\Gamma\left(\frac{Q^2}{\mu^2},\xi,g^2\right)&=&\lim_{\epsilon \rightarrow 0}
Z_\Gamma\left(\frac{1}{\epsilon},\xi,g^2\right)
\Gamma_{\mathrm{B}}\left(Q^2,\xi_{\mathrm{B}},g^2_\mathrm{B},\epsilon\right)\,,\ \ \ \ \
\end{eqnarray}
where $g$ is the gauge coupling, $\xi$ is the gauge fixing parameter, $\mu$ is the renormalization scale, and $\epsilon=2-D/2$.
We have
\begin{equation}
g^2_{\mathrm{B}}=
\mu^{2\epsilon}\left[g^2+\sum_{n=1}^{\infty}a^{(n)}\!\left(g^2\right)\epsilon^{-n}\right],\qquad
\alpha_{\mathrm{B}}=\alpha Z_V\,,\label{gbex}
\end{equation}
where $a^{(n)}$ are polynomials of degree $n+1$ and $Z_V$ is the renormalization constant of the gauge field $V$. 
The background-field method provides a simple connection between the renormalization constants of $g$ and the background gauge field $\hat{V}$,
\begin{equation}
	Z_{g} = Z_{\hat{V}}^{-1/2}\,.
	\label{eq:bftog}
\end{equation}
$Z_\Gamma$ has the following structure:
\begin{eqnarray}
\label{eq:5}
Z_\Gamma\!\left(\frac{1}{\epsilon},\xi,g^2\right)=
1+\sum^\infty_{n=1}c_\Gamma^{(n)}\!\left(\xi,g^2\right)\epsilon^{-n}\,.
\end{eqnarray}
The anomalous dimension of $\Gamma$ is then extracted as  
\begin{eqnarray}
\label{defga}
\gamma_\Gamma(\xi,g^2)=
g^2\frac{\partial}{\partial g^2}\ c^{(1)}_\Gamma\!(\xi,g^2)\,.
\end{eqnarray}

We are thus led to calculate the bare two-point function of the background field $\hat{V}$ at four loops.
To this end, we use \texttt{DIANA} \cite{Tentyukov:1999is}, which generates all the Feynman diagrams with the help of \texttt{qgraf} \cite{Nogueira:1991ex} and calls \texttt{FORM} \cite{Ruijl:2017dtg,Davies:2026cci} to compute the obtained expressions diagram by diagram.
The color algebra is manipulated using \texttt{COLOR} \cite{vanRitbergen:1998pn}, and the Feynman integrals are calculated with \texttt{FORCER} \cite{Ruijl:2017cxj}.

Let us now turn to specific aspects related to DRED.
As mentioned above, despite its self-inconsistency, DRED can still be applied in practice without substantial modifications, provided that the $\epsilon$-scalars are properly taken into account.
Their internal properties should, in particular, coincide with those of the corresponding physical scalars.
This can be understood from the fact that all four-dimensional SYM theories can be viewed as dimensional reductions of $\mathcal{N}=1$ SYM theories formulated in higher (integer) dimensions $\mathcal{D}$ \cite{Gliozzi:1976qd}.
We can then use the following relations in $D=\mathcal{D}-2\epsilon$ dimensions, taken from Ref.~\cite{Avdeev:1980bh} and generalized from their original formulation for $\mathcal{N}=4$ SYM theory to all SYM theories:
\begin{eqnarray}
  \{\alpha^r,\alpha^t\}&=&\{\beta^r,\beta^t\}=-2\delta^{rt}\,,\qquad
  [\alpha^r,\beta^t]=0\,,\nonumber\\
  \alpha^r\alpha^r&=&\beta^r\beta^r=-\delta^{rr}=-n_s-\epsilon\,,\nonumber\\
\mathrm{tr}(\alpha^r\alpha^t)&=&\mathrm{tr}(\beta^r\beta^t)=-n_f\delta^{rt}\,,\qquad 
\mathrm{tr}(\alpha^r\beta^t)=c_{\mathcal{N}}^{rt}\,,\ \ \ \
\label{Relations}
\end{eqnarray}
where $\alpha^r$ and $\beta^r$ comprise the Yukawa couplings, normalized to $g$, of the $n_f$ Majorana spinors to the $n_s=2(n_f-1)$ real scalars and pseudoscalars, respectively.
Furthermore, $c_4^{rt}=0$ \cite{Avdeev:1980bh}, $c_2^{rt}=-2$, and $c_1^{rt}=\mathcal{O}(\epsilon)$ \cite{Velizhanin:2008rw}.
The latter is because the only scalars within $\mathcal{N}=1$ SYM theory are those of $\epsilon$ type.
Notice that $\alpha^r$ and $\beta^t$ are not actually matrices for $\mathcal{N}=1,2$.
Specifically, four-dimensional $\mathcal{N}=1,2,4$ SYM theory, with $n_f=1,2,4$, corresponds to $\mathcal{N}=1$ SYM theory with $\mathcal{D}=n_s+4=2(n_f+1)=4,6,10$.
The advantage of the relations in Eq.~\eqref{Relations} is that they allow one to work with generic value of $\mathcal{N}$ and so to obtain the beta functions all SYM theories in one sweep.
In fact, this is what was originally done at three loops \cite{Avdeev:1982np,Velizhanin:2008rw}.\footnote{%
In the original version of Ref.~\cite{Velizhanin:2008rw}, following Ref.~\cite{Avdeev:1982np}, $c_{\cal N}^{rt}=0$ was wrongly used for $\mathcal{N}=1,2$, too, which resulted in a puzzling non-zero result for the beta function of $\mathcal{N}=2$ SYM theory at three loops, challenging the validity of DRED.}
Recently, an independent three-loop analysis was performed considering the cases $\mathcal{N}=1,2,4$ separately \cite{Chakraborty:2026hdv} and finding agreement with Ref.~\cite{Velizhanin:2008rw}.

With this setup in place, we now proceeded to four loops, similarly as in the SM case \cite{Bezuglov:2026okb}.
In the first step, we compute all the contributing Feynman diagrams using the ``na{\"\i}ve'' prescription for $\gamma_5$, namely, assuming its anticommutation with the other Dirac matrices and nullifying all the traces with odd numbers of $\gamma_5$ inside.
Through four loops, the calculation involves 
63390 diagrams and yields:
\begin{eqnarray}
\lefteqn{\beta_{\mathcal{N}}(a)=\frac{1}{2}(\mathcal{D} - 10)C_A a^2 \left[1 - (\mathcal{D} - 6)C_A a
\vphantom{\frac{7}{4}}\right.}
    \nonumber\\
    &&{}+\left. \frac{7}{4} (\mathcal{D} - 6)^2C_A^2 a^2 - \frac{17}{4} (\mathcal{D} - 6)^3C_A^3 a^3\right]
\label{naivebeta}\\
&&{}-
(\mathcal{D}-4)(\mathcal{D}-10)a^5\left[ C_A^4 \left(\frac{4}{3} - \zeta_3\right) + 4\,d_{44} \big(1 + 6 \zeta_3\big)\right]\,,\nonumber
\end{eqnarray}
where $a=g^2/(4\pi)^2$ and $C_A$ ($d_{44}$) is the quadratic (quartic) Casimir invariant of the adjoint representation.
We observe that the first two lines of Eq.~\eqref{naivebeta} coincide with the known results for $\mathcal{N}=1,4$.
In particular, the beta function for $\mathcal{N}=4$ is found to be zero, as expected \cite{Mandelstam:1982cb,Brink:1982wv,Velizhanin:2010vw,Chakraborty:2026hdv}; this is indicated by the overall factor $(\mathcal{D}-10)$.
In the case of $\mathcal{N}=1$, our four-loop result agrees with Refs.~\cite{Jack:1998uj,Harlander:2006xq,Chakraborty:2026hdv}.
We hence conclude that the ``na{\"\i}ve'' prescription for $\gamma_5$ works perfectly well for $\mathcal{N}=1,4$.
In the case of $\mathcal{N}=2$, we recover the well-known one-loop contribution, but obtain also a non-vanishing term at four loops, as given in the third line of Eq.~\eqref{naivebeta}.

\begin{figure}
\begin{center}
\begin{tabular}{ccc}
\multicolumn{3}{c}{\hspace{-2mm}\includegraphics[width=.49\textwidth]{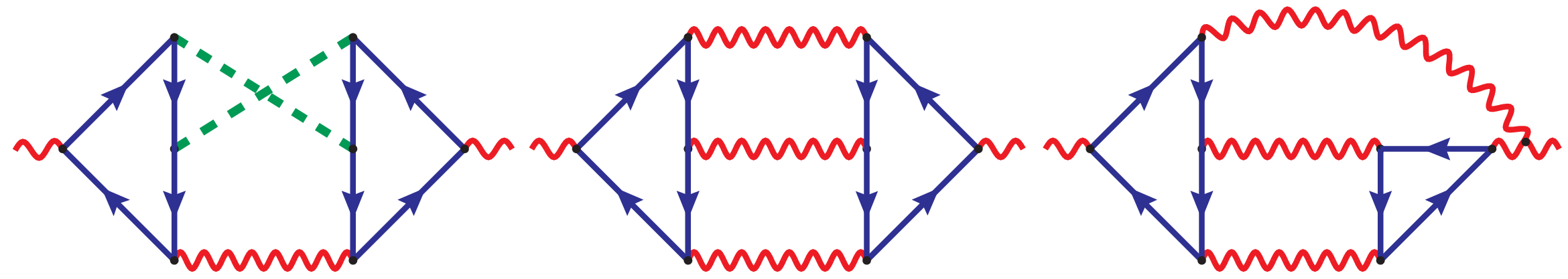}}\\
\hspace {10.5mm}(a) &\hspace {23mm} (b) &\hspace {11mm} (c) 
\end{tabular}
\caption{Typical four-loop Feynman diagrams contributing to the $\hat{V}$ boson self energy with odd numbers of $\gamma_5$ matrices inside fermion traces yielding non-zero contributions.
Diagram~(a) contains four vertices in both fermion loops and two scalar fields (72 diagrams); 
diagram~(b) contains four vertices in both fermion loops and only vector fields (6 diagrams);
and diagram~(c) contains three vertices (triangle) at least in one fermion loop
(68 diagrams).
The diagrams of types (b) and (c) each sum up to zero.
Wavy lines denote vector fields, dashed lines denote scalar fields, and solid lines with arrows denote fermion fields.}
\label{fig:dia}
\end{center}
\end{figure}

In the second step, we re-consider all four-loop Feynman diagrams containing at least two fermion traces.
There are 146 such diagrams; typical examples are depicted in Fig.~\ref{fig:dia}.
We re-compute them using the reading point method \cite{Korner:1991sx}, complemented by the following implementation rules:
(i) re-write all vertices involving fermions with $\omega_\pm=(1\pm\gamma_5)/2$;
(ii) consider only traces that include odd numbers of $\gamma_5$ matrices;
(iii) choose only internal vertices as reading points; and
(iv) sum over all possible choices and divide by their number.
This turns out to be very similar as in the SM case \cite{Bezuglov:2026okb}.
In both cases, we have exactly the same types of Feynman diagrams, but with different internal-group structures.
Thus, the treatment of $\gamma_5$ carries over from the SM to $\mathcal{N}=2$ SYM theory one by one.
We thus obtain precisely the same expression as in the last line of Eq.~(\ref{naivebeta}), but with the opposite sign and multiplied by $(c_{\mathcal{N}}^{rt})^2/4$.
This contribution was wrongly put to zero by the ``na{\"\i}ve'' $\gamma_5$ prescription, so that there is no double counting here.

For $\mathcal{N}=2$, with $c_2^{rt}=-2$, this extra term precisely cancels the disturbing term in the third line of Eq.~\eqref{naivebeta}, in agreement with the respective all-order result \cite{Howe:1983wj}.
For $\mathcal{N}=4$, with $c_4^{rt}=0$, this extra term vanishes, once again in compliance with the respective all-order result \cite{Mandelstam:1982cb,Brink:1982wv}.
For $\mathcal{N}=1$, the four-loop diagrams, stripped bare of the Yukawa couplings, generate a divergence of proportional to $1/\epsilon$, which is compensated by the overall factor $(c_1^{rt})^2$, so that no contribution to the beta function arises.

In conclusion, we have simultaneously subjected two important conceptual elements of DREG \cite{Bollini:1972ui,tHooft:1972tcz}, which have been recurring again and again in the literature over the past fifty years, to a stringent test, namely the consistent treatment of $\gamma_5$ and the protection of supersymmetry.
In fact, we have validated the reading-point method \cite{Korner:1991sx} as implemented in Refs.~\cite{Bednyakov:2015ooa,Bezuglov:2026okb} and the formulation of DRED \cite{Siegel:1979wq} with $\epsilon$-scalars \cite{Capper:1979ns,Nicolai:1980km,Avdeev:1980bh} by ensuring the vanishing of the beta function of $\mathcal{N}=2$ SYM theory beyond one loop, as guaranteed on general grounds \cite{Howe:1983wj}.
Our validation of the extended reading-point methodology \cite{Bednyakov:2015ooa,Bezuglov:2026okb} also carries over to the WCC \cite{Jack:2013sha,Antipin:2013sga,Poole:2019kcm}, which was found to yield identical results for the gauge-coupling beta functions in the SM \cite{Bednyakov:2015ooa,Bezuglov:2026okb}.

\subsection*{Acknowledgments}
This work was supported by the German Research Foundation DFG through Grant No.~
KN~365/16-1.


%

\end{document}